\PassOptionsToPackage{draft}{hyperref}
\documentclass[sigconf,10pt]{acmart}

\usepackage{algorithm}
\usepackage[noend]{algpseudocode}
\usepackage{multirow}
\usepackage{subcaption}

\renewcommand\footnotetextcopyrightpermission[1]{}

\makeatletter
\newcommand{\compactsubsection}{\def\@toclevel{2}%
  \@startsection{subsection}{2}{\z@}%
  {-.5\baselineskip}%
  {.25\baselineskip}%
  {\ACM@NRadjust\@subsecfont}}
\makeatother

\AtBeginDocument{%
  }

\newcommand{\bulletpar}{%
  \par\noindent
  \hangindent=1.2em
  \hangafter=1
  \makebox[1.2em][l]{\textbullet}\ignorespaces}

\setcopyright{none}
\copyrightyear{2018}
\acmYear{2018}
\acmDOI{XXXXXXX.XXXXXXX}
\acmConference[Conference acronym 'XX]{Make sure to enter the correct
  conference title from your rights confirmation email}{June 03--05,
  2018}{Woodstock, NY}
\acmISBN{978-1-4503-XXXX-X/2018/06}

\begin{document}

\title{EStream: Fast and Memory-Efficient MoE Prefill through Expert Virtualization on Mobile NPUs}

\author{Junming Zhang}
\authornote{Junming Zhang and Zhenzhe Zheng contributed equally to this research.}
\email{elliott-z@sjtu.edu.cn}
\orcid{0009-0009-2061-2913}
\affiliation{%
  \institution{Shanghai Jiao Tong University}
  \city{Shanghai}
  \country{China}}

\author{Zhenzhe Zheng}
\authornotemark[1]
\authornote{Corresponding author.}
\email{zhenzhezheng@sjtu.edu.cn}
\affiliation{%
  \institution{Shanghai Jiao Tong University}
  \city{Shanghai}
  \country{China}}

\author{Fan Wu}
\email{wu-fan@sjtu.edu.cn}
\affiliation{%
  \institution{Shanghai Jiao Tong University}
  \city{Shanghai}
  \country{China}}

\author{Xiaoyao Huang}
\email{huangxy32@chinatelecom.cn}
\affiliation{%
  \institution{Cloud Computing Research Institute, China Telecom}
  \country{China}}

\author{Jie Wu}
\email{jiewu@temple.edu}
\affiliation{%
  \institution{Temple University}
  \city{Philadelphia}
  \state{Pennsylvania}
  \country{USA}}

\renewcommand{\shortauthors}{Zhang et al.}

\begin{abstract}
Mobile vendors and application developers increasingly deploy LLMs on
smartphones for diverse prefill-only services. 
Yet current systems rely mainly on dense models
whose regular computation maps efficiently to mobile NPUs, leaving more capable
MoEs underused. MoE prefill does not fit mobile
NPUs: NPU graphs are fixed at compile time, yet MoE picks experts at runtime;
and one request touches most experts, more than a phone can hold in
memory. 
We present \textsc{EStream}, which resolves both by separating what the NPU
must fix from what MoE decides at runtime. A single compiled expert graph
serves every expert, with each expert's routed tokens and weight address bound
at call time, so dynamic MoE execution runs entirely on the NPU without
padding or CPU/GPU fallback. Expert virtualization keeps the expert pool in
UFS flash storage and pages it through a fixed-size NPU-addressable arena,
group by group, with loading hidden behind computation, so memory is bounded
by the arena rather than by the model. It further introduces a hardware-aware
configuration algorithm that automatically configures the UFS--NPU pipeline
and maximizes
loading--computation overlap. Across 18 comparative settings covering three
7B--16B MoEs and 256--4,096-token prompts, we evaluate \textsc{EStream} on a
commercial Snapdragon smartphone. Compared to the fastest baseline at each
setting, \textsc{EStream} achieves a 2.25--27.57$\times$ pure-prefill TTFT
speedup and reduces peak physical memory by 1.19--12.29$\times$.
\textsc{EStream} further scales to MoE models with up to 46.7B parameters.
\end{abstract}

\maketitle
\pagestyle{plain}
\thispagestyle{plain}

\section{Introduction}
\label{sec:introduction}

\begin{figure}[!t]
  \centering
  \includegraphics[width=\columnwidth]{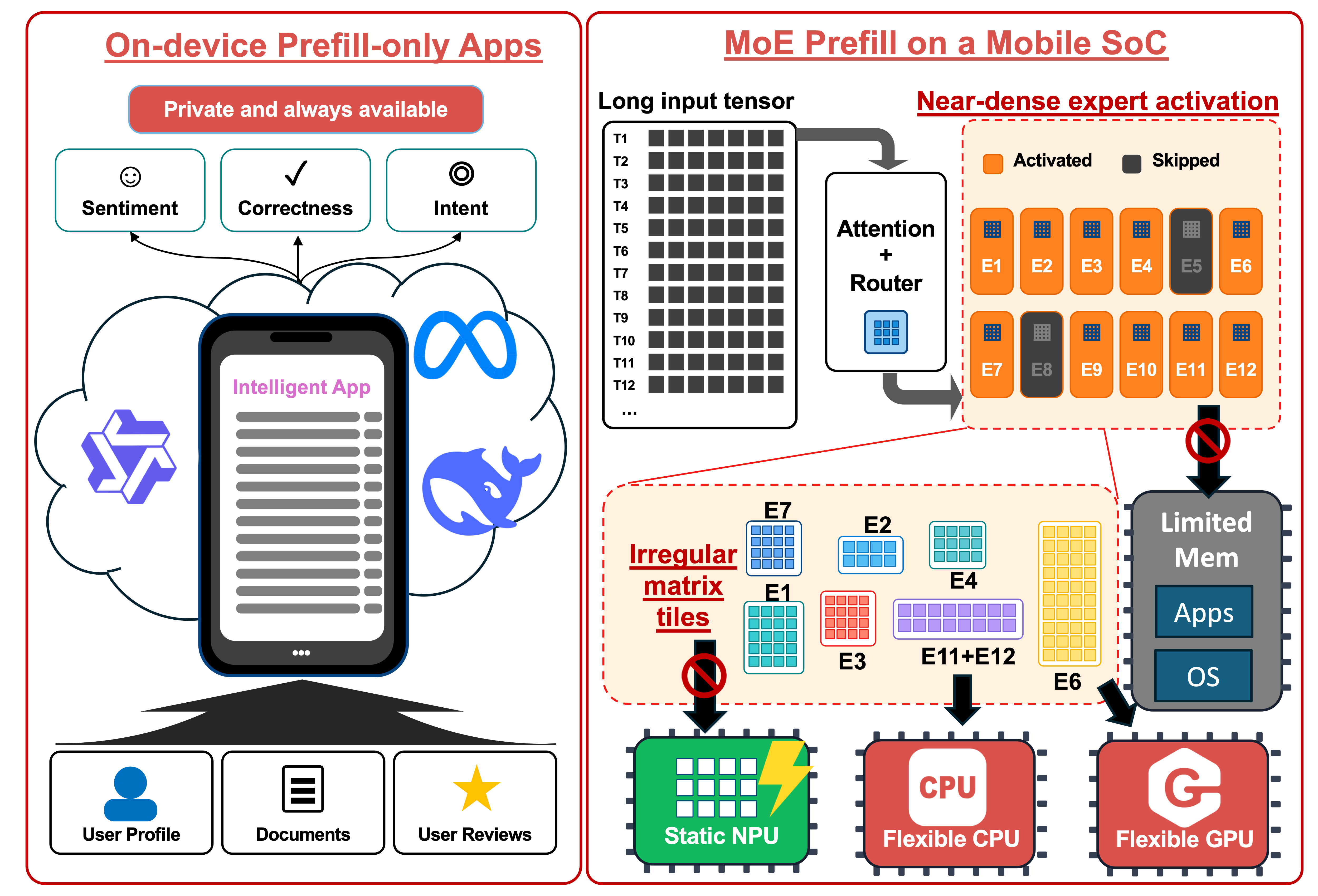}
  \caption{On-device prefill-only applications process private data,
  but MoE prefill remains inefficient on mobile devices.}
  \Description{A two-panel diagram. The left panel shows private user profiles,
  documents, and reviews entering a phone and producing sentiment, correctness,
  and intent decisions. The right panel shows a long tensor routed to nearly
  all experts, irregular matrix tiles that do not map cleanly to a mobile NPU,
  and slower, power-hungry CPU and GPU alternatives.}
  \label{fig:intro-motivation}
\end{figure}

Mobile systems increasingly run language models on the device,
where keeping inference local protects user data and enables operation
under poor network conditions~\cite{zhou2026prism,yin2024llmaas}.
Beyond open-ended content generation, these models serve semantic decision
tasks, including intent recognition, content moderation, recommendation,
factual verification, and candidate ranking~\cite{du2025prefillonly,su2026zeroprefill};
on mobile phones, for example, Android's Gemini Nano flags scam calls entirely on the
device~\cite{google2025scamdetection}, and Apple's assistant maps requests to
intents~\cite{aas2023ondevice} (Figure~\ref{fig:intro-motivation}, left).
These workloads obtain their useful outputs from prefill-stage logits without
iterative autoregressive decoding, are known as
\emph{prefill-only workloads}, and already constitute a substantial class of
user requests~\cite{su2026zeroprefill}.

However, existing mobile systems remain designed primarily
around dense Transformers~\cite{zhou2026prism,chen2026mobilemoe}, creating a  mismatch with the broader LLM landscape, where sparse Mixture-of-Experts (MoE) 
has emerged as a leading architecture for scaling model capability without a proportional
increase in per-token computation~\cite{fedus2022switch,du2022glam,dai2024deepseekmoe}.
MoE can outperform dense models under similar inference-FLOP budgets
~\cite{dai2024deepseekmoe,muennighoff2025olmoe,qwen2025qwen3,
chen2026mobilemoe}.
For example,
Qwen3-30B-A3B holds 30B parameters but activates only 3B per token, yet outperforms dense Qwen3-4B across all reported
benchmarks~\cite{qwen2025qwen3}.
This capacity--computation advantage could enable more capable
mobile services, such as on-device content moderation and personalized
content recommendation~\cite{firooz2025brew}.
Apple's newly announced on-device model holds 20B sparse parameters
in flash, yet routes experts per prompt rather than per token because
per-token expert swapping is too slow~\cite{apple2026afm3}.

Realizing this opportunity on phone-class SoCs remains difficult because MoE
prefill creates two coupled mismatches. \textbf{First, dynamic MoE execution conflicts with graph-centric NPU interfaces.}
Each layer reveals
which experts are active, and how many tokens each receives, only after routing. Although low-level NPU
primitives support runtime control, addressing, and data movement
~\cite{hao2026mobilenpu}, common deployment stacks expose the NPU
only through precompiled graphs with fixed tensor shapes
~\cite{xu2025llmnpu,chen2025heteroinfer}. Existing NPU systems therefore
compile graphs of fixed capacity and invoke them repeatedly until all
routed tokens are processed, while retaining routing,
indexing, or dispatch on the CPU/GPU~\cite{benazir2026npumoe}
(Figure~\ref{fig:intro-motivation}, right, bottom). This adds
padding, host-control, and cross-processor synchronization overheads.
\textbf{Second, token-level sparsity becomes nearly dense at request level.}
Although each token activates only a few experts, multi-token prefill
collectively touches most of the expert pool
(Figure~\ref{fig:intro-motivation}, right, top), causing severe memory pressure in
resident runtimes~\cite{su2026zeroprefill}.
Existing remedies relieve this pressure at a cost.
CPU/GPU offloading systems
handle dynamic routing and reduce expert residency by caching and
prefetching, but do so on the CPU/GPU rather than the NPU
~\cite{xue2024moeinfinity,yi2025edgemoe,yang2026zipmoe}. Model-side pruning or
expert substitution trades accuracy for efficiency
~\cite{lu2024notall,cheng2025hookmoe,cao2026condense,hao2026lightmoe}.

In this work, we present \textsc{EStream}, a low-latency,
memory-efficient, and accuracy-preserving system for on-device MoE prefill.
First, for the graph mismatch, \emph{topology-invariant expert graph sharing}
compiles one expert graph for all experts of the same shape, and
\emph{runtime route and parameter binding} passes each expert's tokens and
weight address as inputs on every call; with routing also on the NPU, no
expert is padded and no step returns to the CPU/GPU, i.e., \emph{full-NPU}
prefill. Second, for the memory
mismatch, \emph{expert virtualization} keeps the expert pool in UFS flash
storage and streams experts, a group at a time, through a fixed-size arena
the NPU reads directly, freeing each group's space after its last use; a
request may activate every expert, yet only one arena's worth is ever in
memory. Third, to hide the storage latency that this streaming introduces, a
\emph{UFS--NPU pipeline} overlaps each group's load with non-expert
computation and execution of groups already loaded, with group size, queue
depth, and resident group count automatically configured.

We implement \textsc{EStream} on a OnePlus 15 with a Snapdragon 8 Elite Gen~5
SoC using weight-only Q4 quantization, floating-point activations, and no
approximation-based sparsity. We evaluate OLMoE, LFM2.5, and DeepSeek-V2 across
18 model--length settings up to 4,096 tokens, with additional experiments on
Qwen3-30B-A3B and Mixtral-8$\times$7B.

This paper makes the following contributions:

\bulletpar We identify two mismatches that block full-NPU MoE prefill on mobile
devices. Input-dependent expert execution conflicts with graph-centric NPU
stacks that bind graph structure, tensor capacity, and parameters ahead of
execution, while request-level densification activates nearly the entire
expert pool even when device memory can hold only a fraction of it.

\bulletpar We design and implement \textsc{EStream}, which reuses a
topology-invariant graph across shape-compatible experts, binds routes and
parameters at runtime, virtualizes the expert pool over a bounded
NPU-addressable arena streamed from UFS, and automatically configures the
UFS--NPU pipeline. It thereby enables fast and memory-efficient MoE prefill
on mobile SoCs without modifying the model. To our knowledge, it is the first
system to run full-NPU large MoE prefill on a commercial
smartphone.

\bulletpar Across the 18 primary model--length settings, \textsc{EStream}
achieves a 2.25--27.57$\times$ pure-prefill TTFT speedup and uses
6.45--12.29$\times$ less physical memory than the fastest completed
non-offloading baseline at each point. Against a same-backend reference with
every expert resident, it uses 3.73--6.36$\times$ less memory, while its
latency gap narrows to 4.3--23.0\% at 4K.

\section{Background And Motivation}
\label{sec:background}

\subsection{NPUs on Mobile SoCs}
\label{sec:background-mobile-soc}

As on-device AI has advanced, hardware vendors have added neural processing
units (NPUs) to unified-memory mobile SoCs, creating heterogeneous systems in
which CPUs, GPUs, and NPUs coexist and share physical DRAM. Mobile applications
typically access the NPU through a vendor graph runtime. On Qualcomm platforms,
QNN represents a neural network as a graph with predefined operator topology,
tensor capacities, and parameter bindings, then finalizes it into an executable
context. The underlying Hexagon Tensor Processor (HTP) is more flexible than
this abstraction suggests. It combines scalar control threads, wide SIMD vector
units (HVX), matrix engines (HMX), DMA engines, and software-managed on-chip
memory (VTCM)~\cite{qualcomm2021hvx,qualcomm2026hmx}. HMX performs tiled matrix
multiplication, HVX handles vector operations and data rearrangement, scalar
threads resolve runtime control and addresses, and DMA moves tensor tiles
between DRAM and VTCM. These resources can process different tiles
concurrently, forming an internally heterogeneous NPU pipeline. This
organization is particularly effective for MoE prefill, where routed tokens
form large matrix multiplications. Figure~\ref{fig:expert-npu-opportunity}
shows that experts with at least 64 routes account for 94.8\% of CPU expert
projection time during a 1K-token OLMoE prefill. At this route count, the NPU
is 4.4$\times$ and 4.6$\times$ faster than the evaluated CPU and GPU.

\subsection{Obstacles to On-Device MoE Prefill}
\label{sec:background-prefill}

A prefill-only request processes $T$ prompt tokens in one causal
forward pass and derives its result from the final hidden state or next-token
distribution, without autoregressive decoding
~\cite{du2025prefillonly,su2026zeroprefill}. Its on-device objectives are low
latency and a small memory footprint. In an
MoE layer, a router assigns each token to the top-$k$ of $E$ experts, increasing
model capacity without proportionally increasing computation
~\cite{fedus2022switch,dai2024deepseekmoe}. Unlike dense models, whose regular
computation maps readily onto mobile-SoC NPUs, MoE's prefilling creates
two obstacles to efficient on-device deployment.

\begin{figure}[!t]
  \centering
  \includegraphics[width=0.875\columnwidth]{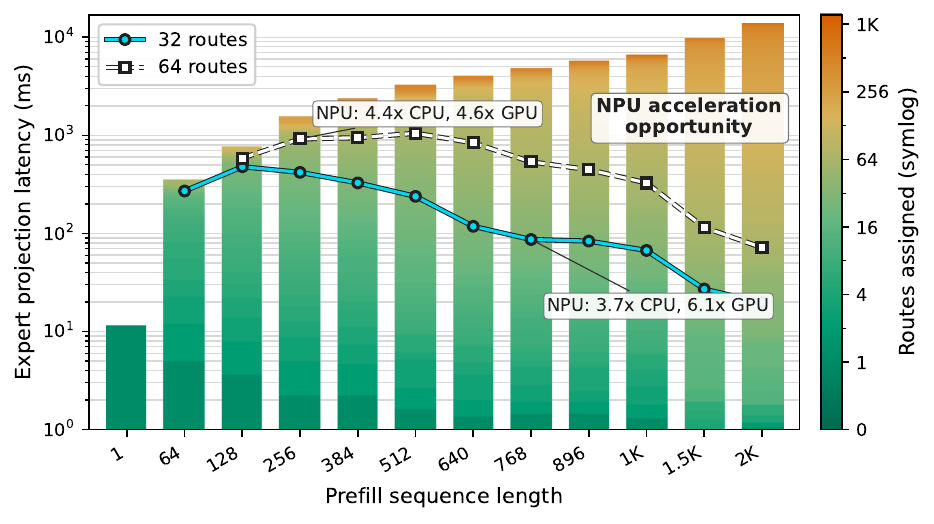}
  \Description{Stacked per-expert projection latency for OLMoE CPU prefill from
  1 to 2048 tokens. Each bar stacks experts in increasing routed-token count,
  with color indicating route count. Curves mark the boundaries above which
  experts receive at least 32 or 64 routes. The region above these curves grows
  with prompt length and is labeled as an NPU acceleration opportunity.}
  \caption{Longer prompts expose more expert computation to NPU acceleration.
  Each bar stacks the execution latencies of all experts during OLMoE CPU
  prefill at one input length.}
  \label{fig:expert-npu-opportunity}
\end{figure}

\noindent\textbf{Input-dependent expert execution.}
Active experts and each expert's routed-token count $r_e$ become known only
after routing the layer input. Although the expert operator sequence
remains fixed, each request produces a different set of expert invocations,
gathered tensors of extent $r_e\times d$, and selected parameter tensors. This
execution pattern conflicts with graph-centric NPU interfaces that optimize around
finalized contexts with predetermined graph structure, capacity bounds, and
parameter bindings~\cite{xu2025llmnpu,chen2025heteroinfer}. Padding every expert
to a worst-case capacity wastes computation, while switching among shape- or
expert-specific contexts adds memory, launch, and synchronization overheads.
CPU/GPU orchestration avoids these restrictions but leaves the NPU acceleration
opportunity in Figure~\ref{fig:expert-npu-opportunity} unused.

\begin{figure}[t]
  \centering
  \begin{minipage}[t]{0.485\linewidth}
    \centering
    \includegraphics[width=\linewidth]{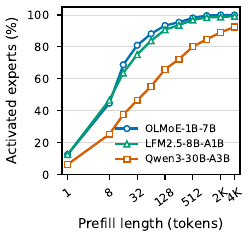}

    \scriptsize\textbf{(a)} Expert activation
  \end{minipage}
  \hfill
  \begin{minipage}[t]{0.485\linewidth}
    \centering
    \includegraphics[width=\linewidth]{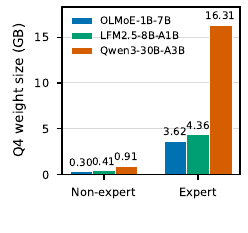}

    \scriptsize\textbf{(b)} Q4 weight distribution
  \end{minipage}
  \Description{Two plots characterize the expanded MoE parameter working set.
  Panel (a) shows the percentage of layer--expert pairs activated by C4 prefixes
  from 1 to 4096 tokens; OLMoE and LFM2.5 approach 100 percent, while Qwen3
  reaches 92.4 percent. Panel (b) compares the Q4 payload of non-expert and
  expert weights; expert weights account for more than 91 percent in all three
  models.}
  \caption{MoE prefill activates most experts, which dominate the model
  weights. (a) Activated layer--expert pairs. (b) Q4 expert and non-expert
  weight sizes.}
  \label{fig:moe-prefill-motivation}
\end{figure}

\noindent\textbf{Request-level expert densification.}
Although each token activates only a few experts, their union across
multi-token prefill can cover nearly the entire expert pool.
Figure~\ref{fig:moe-prefill-motivation}(a) shows that 512-token prompts activate
80.1--98.1\% of layer--expert pairs in our evaluated models, increasing to
92.4--100\% at 4K tokens. Expert parameters account for 91.4--94.7\% of
their whole weight payloads, thus leading to high memory pressure
(Figure~\ref{fig:moe-prefill-motivation}(b)).
Demand-paged \texttt{mmap} delays rather than eliminates physical residency.
Executing an expert touches its weight pages and faults them into memory. These
pages accumulate when memory is available, while memory pressure introduces
reclamation and subsequent refaults on the execution path, 
which can stall execution, impair system responsiveness, and degrade user
experience.

\subsection{Opportunities and Challenges}
\label{sec:background-opportunities}

\noindent\textbf{Observation 1: Expert-topology invariance.}
Experts with compatible dimensions perform the same sequence of gather,
gate/up projection, activation, down projection, and weighted scatter-add.
Only their routed tokens, coefficients, and parameters change at runtime. 
This invariance creates an opportunity to keep one
expert graph resident and bind it to request-specific data. Conventional NPU
contexts, however, couple graph topology with tensor shapes and parameters.
Parameter-update interfaces are narrowly supported and still require host calls and
synchronization. The challenge is to dynamically vary routes and parameter addresses
without rebuilding or reloading expert-specific NPU contexts.

\noindent\textbf{Observation 2: Logical activation does not require physical co-residency.}
Request-level densification determines which experts are touched, but their
parameters need not coexist in memory. Expert weights are immutable, stateless,
and share a uniform layout. The NPU consumes them group by group, and a resident
copy is no longer live once its final routed output has been accumulated. This observation
therefore allows the runtime to reuse a bounded NPU-addressable memory region 
across expert groups. The challenge is to make this memory reusing lightweight and
transparent to the reusable NPU graph. Parameters must be
updated correctly and safely without intermediate copy or interruption.

\noindent\textbf{Observation 3: Compute and expert I/O scale differently with prompt length.}
NPU work grows with prompt length, whereas expert I/O approaches the fixed size
of the expert pool once routing densifies. Longer prompts thus provide more
computation behind which UFS reads can be hidden. Realizing this opportunity
requires saturating UFS bandwidth and coordinating a bounded parameter arena
with NPU execution. The runtime must size the arena, issue reads early enough
to hide overhead, and synchronize the I/O producer with the NPU consumer before
parameters are published or reused. To reduce the overall latency, I/O concurrency, transfer granularity,
arena capacity, and loading schedule must maximize I/O--execution overlap.

\section{EStream System Design}
\label{sec:design}

\subsection{System Overview}
\label{sec:design-overview}

\begin{figure}[!t]
  \centering
  \includegraphics[width=\columnwidth]{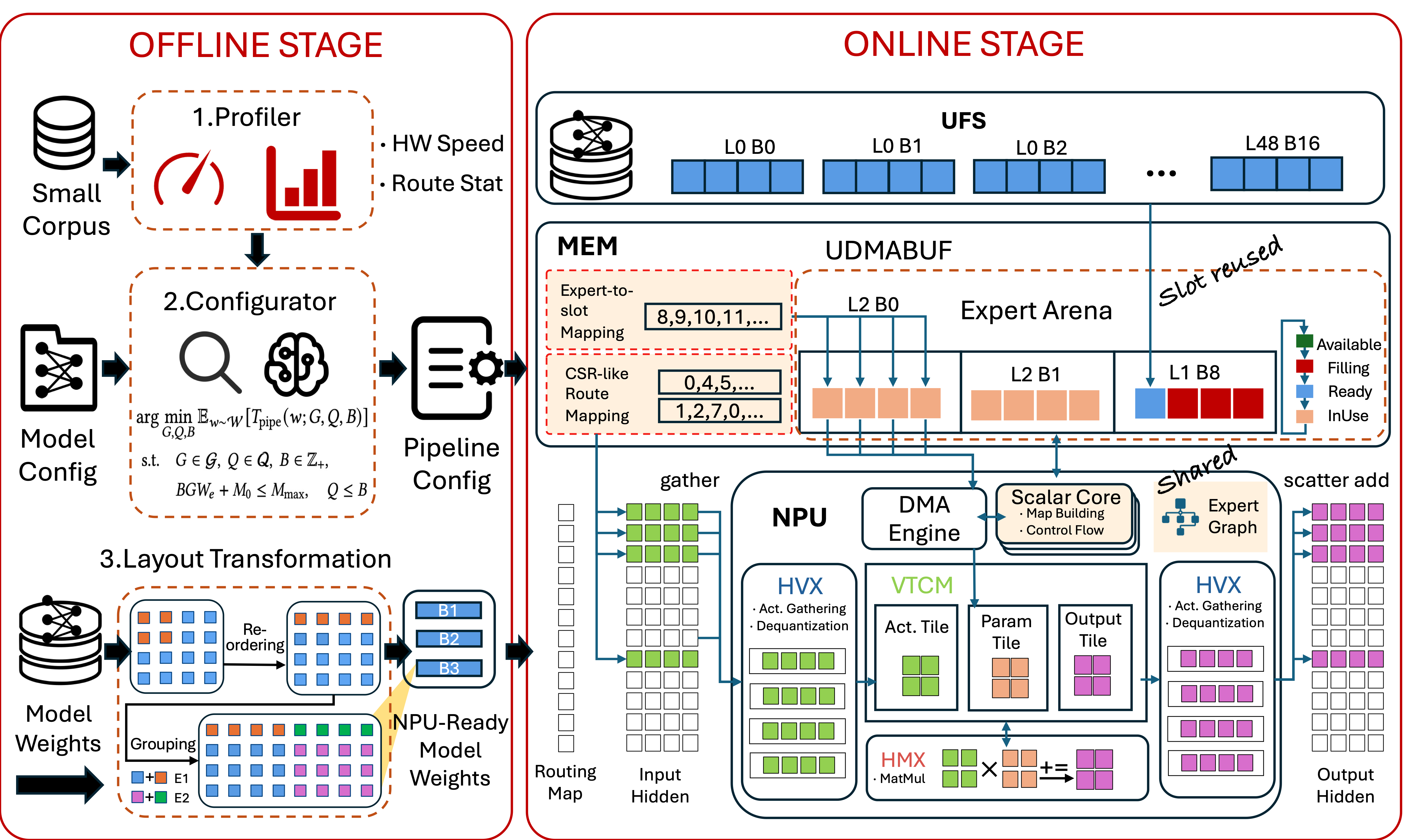}
  \caption{The system overview of EStream.}
  \Description{A two-panel diagram. The left panel shows private user profiles,
  documents, and reviews entering a phone and producing sentiment, correctness,
  and intent decisions. The right panel shows a long tensor routed to nearly
  all experts, irregular matrix tiles that do not map cleanly to a mobile NPU,
  and slower, power-hungry CPU and GPU alternatives.}
  \label{fig:overview}
\end{figure}

Figure~\ref{fig:overview} presents the overview of \textsc{EStream}.
\textsc{EStream} decouples reusable NPU computation graphs from
request-dependent inputs and expert parameters, and organizes the resulting
operations into an efficient NPU compute pipeline. It further virtualizes the
experts over a bounded NPU-addressable memory arena, enabling
the prefill of large MoE models without keeping all experts resident. A coordinated
UFS--NPU pipeline overlaps expert streaming with computation to reduce the
latency introduced by bounded residency.

\noindent\textbf{Offline preparation.}
\textsc{EStream} constructs reusable NPU graphs for the non-expert and expert
computation, and converts expert parameters into the layout consumed directly
by the NPU. It groups the packed experts for efficient storage access and
profiles UFS and NPU service times to select the group size, I/O queue depth,
and number of resident banks under a given memory budget.

\noindent\textbf{Online execution.}
At initialization, \textsc{EStream} creates the reusable graphs and allocates
the bounded expert arena. For each request, the host streams expert groups into
free banks and publishes their runtime bindings, while the NPU executes
non-expert or already available expert computation. The NPU graph consumes 
the dynamic routes and parameter bindings without reconstruction. Once an
expert group has finished execution, its bank is reclaimed for subsequent
experts, forming a continuous loading--execution pipeline.

\subsection{Intra-NPU MoE Execution}
\label{sec:design-in-htp}

MoE execution contains state that changes at three timescales. The operator
topology and its dependencies remain fixed within a compatible expert shape
class. Each invocation determines the selected experts, routed token rows, and
parameter locations. During that invocation, a tile scheduler assigns work to
the heterogeneous resources inside the NPU. \textsc{EStream} represents these
three levels independently. It constructs the expert topology once, supplies
routing and parameter placement as invocation-time metadata, and schedules the
resulting route blocks at tile granularity. One topology can therefore serve
changing expert parameters and route extents.

\noindent\textbf{Topology-Invariant Expert Graph Sharing.}
For each compatible shape class, \textsc{EStream} constructs one shared
routed-expert graph. For expert $e$, let $W_e^{g}$, $W_e^{u}$, and $W_e^{d}$
denote its gate, up, and down parameters. Given token representation $X_t$,
the expert function is
\begin{equation}
F_e(X_t)=W_e^{d}\!\left(
  \operatorname{SiLU}(W_e^{g}X_t)\odot W_e^{u}X_t
\right).
\label{eq:expert-function}
\end{equation}
If $\mathcal R_l(t)$ contains the expert--router-weight pairs selected for
token $t$ in sparse layer $l$, the layer computes
\begin{equation}
Y_t=A_t+\sum_{(e,\alpha)\in\mathcal R_l(t)}\alpha F_e(X_t),
\label{eq:routed-expert-accumulation}
\end{equation}
where $A_t$ is the incoming residual. This defines a fixed dataflow:
gather routed rows, evaluate gate and up projections, apply SwiGLU, evaluate
the down projection, and scatter-add the weighted outputs.

Experts in the same shape class share hidden and feed-forward dimensions,
parameter types and layouts, and operator dependencies. Expert identity
changes only the parameter tensors, while routing changes only token rows and
coefficients. \textsc{EStream} supplies both as invocation-time metadata,
allowing one graph to serve all matching experts and sparse layers without
reconstruction.

\begin{figure}[t]
  \centering
  \begin{subfigure}[t]{0.485\columnwidth}
    \vspace{0pt}
    \centering
    \begin{minipage}[c][0.9\linewidth][c]{\linewidth}
      \centering
      \includegraphics[width=\linewidth]{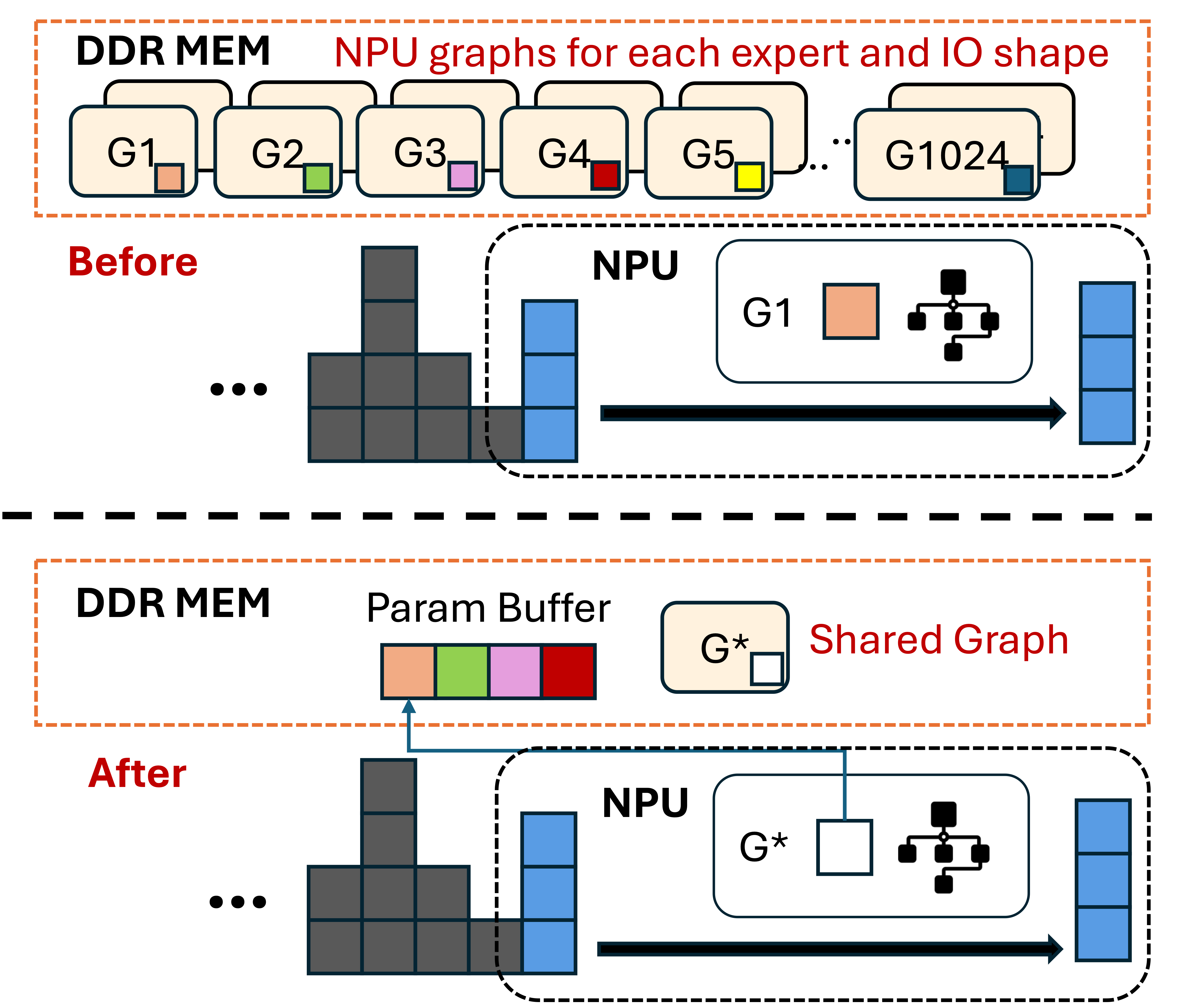}
    \end{minipage}
    \caption{Shared expert graph.}
    \label{fig:shared-expert-graph}
  \end{subfigure}
  \hfill
  \begin{subfigure}[t]{0.485\columnwidth}
    \vspace{0pt}
    \centering
    \begin{minipage}[c][0.9\linewidth][c]{\linewidth}
      \centering
      \includegraphics[width=0.9\linewidth]{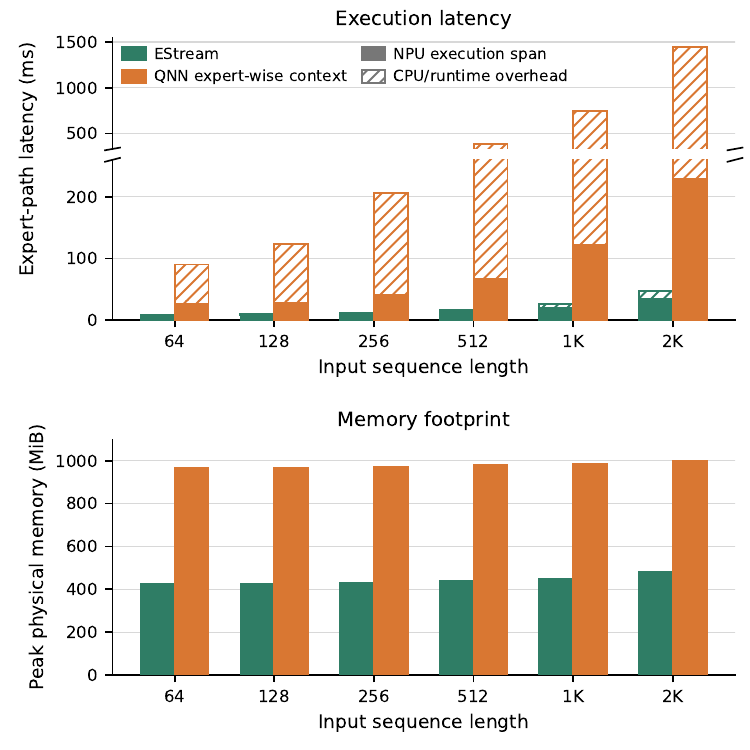}
    \end{minipage}
    \caption{Resident-path comparison.}
    \label{fig:shared-expert-context-comparison}
  \end{subfigure}
  \caption{Topology-invariant graph sharing improves efficiency. (a) A shared
  graph binds data at runtime. (b) Latency and memory versus per-expert QNN
  contexts.}
  \Description{Two subfigures. The left compares parameter-bound expert graphs
  with one shared graph whose parameter buffer is selected at runtime. The
  right uses grouped bars to compare EStream and per-expert QNN contexts in
  expert-path latency and peak physical memory from 64 to 2048 input tokens.}
  \label{fig:graph-sharing-design-results}
\end{figure}

As shown in Figure~\ref{fig:shared-expert-graph}, the shared graph allocates
stable input, accumulator, descriptor, and expert-arena buffers during
initialization. Its descriptors reserve capacity for at most $C$ routes and
$T_{\max}$ source tokens, but execution uses the exact runtime extent: an
invocation with $R\leq C$ evaluates exactly $R$ routes, while larger route
sets are divided among repeated invocations. Parameter views refer to arena
slots whose contents may change after prior uses complete. Switching experts
or entering another compatible sparse layer therefore requires only new route
descriptors and parameter bindings, not graph reconstruction or tensor
reallocation.

The same reusable context hosts the fused expert operator described below.
The next part explains how each invocation constructs its route and parameter
bindings.

\noindent\textbf{Composed Sparse Route and Parameter Binding.}
A shared graph becomes dynamic through two runtime mappings. For invocation
$q$ of sparse layer $l$, the router supplies
$\mathcal{E}=(e_i)_{i=0}^{R-1}$, $\mathcal{T}=(t_i)_{i=0}^{R-1}$, and
$\mathcal{A}=(\alpha_i)_{i=0}^{R-1}$, where route $i$ sends token $t_i$ to
logical expert $e_i$ with weight $\alpha_i$. The host stream manager maintains
a placement map $\Phi_l$: $\Phi_l(e)=s$ means that expert $e$ occupies physical
slot $s$, whereas $\Phi_l(e)=\bot$ means that it is not resident. A binding is
published only after the gate, up, and down slices of the slot have been
populated, and an invocation is submitted only when every referenced binding
is valid.

The NPU composes the two mappings by first resolving each route's physical slot
$\psi_i=\Phi_l(e_i)$. Scalar workers count routes with the same $\psi_i$ and
use prefix sums of these counts to determine where each slot begins. A second
pass stores $(t_i,i)$ contiguously by slot. The records for slot $s$ occupy
\[
\mathit{rec}[\mathit{offset}[s]{:}\mathit{offset}[s+1]).
\]
The retained route index identifies $\alpha_i$ during accumulation. This
compressed sparse row-style representation requires $O(P+R)$ time and space,
skips empty slots, and is rebuilt as routing and residency change.

The graph retains fixed tensor bases for the source activation $X$, destination
accumulator $Y$, and three parameter planes
$\mathcal{B}^{g}$, $\mathcal{B}^{u}$, and $\mathcal{B}^{d}$. Let $s$
be one physical slot from each plane. The address is
\begin{equation}
p_s^j=p_{\mathcal{B}^j}+sS_j,
\qquad j\in\{g,u,d\},
\end{equation}
where $S_j$ is the prepacked size of projection $j$. For each slot,
\textsc{EStream} joins $(p_s^g,p_s^u,p_s^d)$ with every $(t_i,i)$ in that
slot's route segment. The resulting records contain the token row, route weight
$\alpha_i$, and parameter addresses. They select activation row
$p_X+t_iS_X$ and accumulation row $p_Y+t_iS_Y$. NPU vector units gather the
selected rows, DMA transfers parameter tiles, and matrix units evaluate the
projections; weighted outputs are scatter-added to their original positions.

Only the route descriptors, valid extent, and placement map change across
invocations. The graph topology, tensor bases, and strides remain fixed,
allowing different expert invocations to reuse one graph without graph
reconstruction or tensor reallocation.
Algorithm~\ref{alg:shared-expert-execution} summarizes the composed bindings
and their subsequent tile-level execution.

\noindent\textbf{Fused Expert Execution with Intra-NPU Pipelining.}
A routed expert comprises activation gather, gate/up projections, SwiGLU, down
projection, and weighted scatter-add. Separate NPU operators introduce
repeated dispatch boundaries, materialize intermediates in shared DRAM, and
prevent overlap among stages using different resources. \textsc{EStream}
recognizes this complete pattern during graph construction and replaces it
with one route-aware fused operator that consumes the upstream sparse route
descriptors.

VTCM capacity determines the operator's tiling and concurrency.
\textsc{EStream} selects route-row and projection-column tile sizes that fit
the available capacity, then partitions VTCM by data lifetime. One region
holds gathered activations, while another retains completed SwiGLU tiles until
the down projection consumes them. Two compressed-weight buffers receive
alternating DMA transfers, and two decoded-weight buffers let vector workers
prepare the next matrix operand while the matrix engine consumes the current
one. Four output buffers retain two adjacent gate/up tile pairs, allowing the
matrix engine to produce the next pair while vector workers apply SwiGLU to
the previous pair. The down projection reuses two output buffers to overlap
matrix computation with weighted scatter-add. This lifetime-aware layout
prevents concurrent stages from overwriting live values without spilling
intermediate tiles to shared DRAM.

\begin{algorithm}[t]
\caption{Composed bindings and VTCM tile pipeline}
\Description{Pseudocode for binding routed tokens and resident parameters to a
shared graph, retaining intermediate tiles in VTCM, and overlapping DMA, vector,
and matrix work across tiles.}
\label{alg:shared-expert-execution}
\fontsize{7.5pt}{8.5pt}\selectfont
\begin{algorithmic}[1]
\Require Routes $\rho=\{(e_i,t_i,\alpha_i)\}_{i=0}^{R-1}$ and placement $\Phi_l$
\Require Arena $\mathcal B=(\mathcal B^g,\mathcal B^u,\mathcal B^d)$
\Require tensors $(X,Y)$ and VTCM budget $V$
\State $\psi_i\gets\Phi_l(e_i)$ for $i=0,\ldots,R-1$
       \Comment{Resolve physical slots}
\State $\mathit{count}[s]\gets\sum_i[\psi_i=s]$
\State $\mathit{offset}[0]\gets0$;
       $\mathit{offset}[s{+}1]\gets\mathit{offset}[s]+\mathit{count}[s]$
       for $s=0,\ldots,P-1$ \Comment{Prefix-sum offsets}
\State $\mathit{cursor}\gets\mathit{offset}$
\For{$i\gets0$ to $R-1$}
  \State $p\gets\mathit{cursor}[\psi_i]$
  \State $\mathit{rec}[p]\gets(t_i,i)$;
         $\mathit{cursor}[\psi_i]\gets p+1$
         \Comment{Build route map}
\EndFor
\State $\mathcal V\gets\Call{PlanVTCM}{V;Z,H,C[2],D[2],O[4]}$
       \Comment{No DDR spills}
\ForAll{$s$ such that $\mathit{count}[s]>0$}
  \State $\mathcal W_s^j\gets p_{\mathcal B^j}+sS_j$,
         $j\in\{g,u,d\}$
         \Comment{Build parameter map}
  \State $\mathcal Q_s\gets
         \mathit{rec}[\mathit{offset}[s]{:}\mathit{offset}[s+1])$
  \State $\mathcal D_s\gets\Call{AttachWeights}{\mathcal Q_s,\rho,\mathcal W_s}$
         \Comment{Compose both maps}
  \ForAll{$Q$ in \Call{Chunks}{$\mathcal D_s$}}
    \State $Z\gets\Call{GatherToVTCM}{X,Q,\mathcal V.Z}$
    \State $\mathcal J\gets\Call{GateUpTiles}{\mathcal W_s^g,\mathcal W_s^u}$
    \For{$j$ in \Call{PipelineSteps}{$\mathcal J$}}
      \State $\mathrm{DMA}(j{+}2)\parallel
        \mathrm{HVX}(\mathrm{dec}_{j+1},\mathrm{act}_{j-2})\parallel
        \mathrm{HMX}(\mathrm{gemm}_{j})$
    \EndFor
    \State $\mathcal K\gets\Call{DownTiles}{\mathcal W_s^d}$
    \For{$c$ in \Call{PipelineSteps}{$\mathcal K$}}
      \State $\mathrm{DMA}(c{+}2)\parallel
        \mathrm{HVX}(\mathrm{dec}_{c+1},\mathrm{scatter}_{c-1}{\to}Y)\parallel
        \mathrm{HMX}(\mathrm{gemm}_{c})$
    \EndFor
  \EndFor
\EndFor
\State \Return $Y$
\end{algorithmic}
\end{algorithm}

Using this layout, DMA fetches future prepacked parameter tiles, vector workers
decode them into matrix-engine operands, and the matrix engine evaluates the
current tile. Gate/up tiles are issued as pairs and transformed in VTCM before
feeding the down projection. Completed down tiles are multiplied by router
weights and scatter-added directly into the final token accumulator. This
DMA--vector--matrix pipeline overlaps parameter preparation, projection, and
postprocessing within one NPU invocation, reducing dispatch overhead and
shared-DRAM traffic without materializing dense per-expert outputs.

\subsection{Expert Virtualization with UFS Streaming}
\label{sec:design-parameter-stream}

The preceding subsection allows one expert topology to consume parameters from
runtime-selected addresses. \textsc{EStream} uses this capability to bound the
memory cost of request-level expert densification. Rather than retaining every
expert that a request may eventually activate, it maintains a fixed
NPU-addressable arena, moves groups of expert records through that arena, and
stores those records in the layout required for immediate NPU execution. This
section describes the arena, the grouping granularity, and the corresponding
offline storage layout in turn.

\noindent\textbf{Bounded Parameter Arena and Safe Slot Reuse.}
Expert parameters are read-only and stateless. Physical slots can therefore be
reused, but reuse separates each expert's stable logical identity from its
changing physical location.

\textsc{EStream} implements the placement map as a stateful expert residency
table. Each valid entry records the expert's current slot $\Phi_l(e)$ and that
slot's state, while a nonresident expert has no valid entry. Like a
software-managed page table, it translates logical expert IDs into physical
locations and indicates whether each mapping is safe to consume. The scheduler
uses only ready entries and passes their slot indices to the shared NPU graph.

The bounded arena contains $P$ fixed-address slots, each holding the gate, up,
and down parameters of one expert. Their graph-visible addresses and strides
remain unchanged. The host updates only the slot contents and residency table.
For a packed expert size $W_e$, these slots consume
$M_{\mathrm{arena}}=P W_e$, so expert residency is independent of the number
of logically activated experts.

An \underline{Available} slot is reserved for an expert before loading, at
which point its new owner is recorded and the slot becomes
\underline{Filling} while its mapping remains hidden. After all parameter
slices arrive, the host publishes the slot as \underline{Ready}. NPU
submission then marks it \underline{InUse}, and the final completion event
invalidates the binding and returns the slot to the \underline{Available}
pool. A \underline{Ready} slot with no routed work may be released directly.

This protocol exposes no partially loaded expert and prevents reassignment
while the NPU may still access a slot. It therefore enables bounded memory
reuse without graph reconstruction or changes to model semantics.

\noindent\textbf{Expert Grouping and Banked Residency.}
Loading experts individually creates short UFS requests that repeatedly incur
I/O submission, completion, and NPU invocation overheads. \textsc{EStream}
amortizes these fixed costs by transferring $G$ experts as one storage group.
An offline permutation $\pi_l(e)$ determines the storage order of experts in
layer $l$, with every $G$ consecutive positions forming one group. Each physical
bank mirrors this organization with $G$ slots in the bounded arena, so loading
one storage group populates one bank. Grouping changes only the granularity of
transfer and residency. The fused operator still evaluates only experts with
nonempty routes.

For expert $e$, let $q_l(e)$ denote its storage group and $r_l(e)$ its position
within that group. If group $q_l(e)$ resides in bank $b$, then
\begin{equation}
\begin{aligned}
q_l(e)&=\left\lfloor\frac{\pi_l(e)}{G}\right\rfloor,\qquad
r_l(e)=\pi_l(e)\bmod G,\\
\Phi_l(e)&=bG+r_l(e).
\end{aligned}
\label{eq:group-slot-binding}
\end{equation}
The stateful residency table publishes these translations only after the entire
bank becomes ready and invalidates them together when the bank is recycled.

With $B$ resident banks, the arena contains $P=BG$ slots and bounds expert
residency at $M_{\mathrm{arena}}=BGW_e$. A larger $G$ amortizes fixed transfer
and dispatch costs, but may fetch inactive experts and leaves fewer groups
available for pipeline overlap. A smaller $G$ offers finer scheduling
granularity at the cost of fragmented I/O and more runtime operations. More
banks permit deeper read-ahead while increasing memory proportionally.
Section~\ref{sec:design-pipeline} selects $G$ and $B$ jointly under the
deployment memory budget.

\noindent\textbf{Offline NPU-Consumable Weight Layout.}
Serialized model-weight files typically
arrange each matrix in a storage-oriented, row-major layout with
format-specific packing. In contrast, the NPU operator consumes $32\times32$
tiles along the reduction dimension and expects values from multiple output
rows in its tile traversal order. Loading the serialized bytes unchanged would
require the host to rearrange every expert before publishing its
bank, adding temporary buffers, memory copies, and DRAM traffic to the critical
path.
To remove this latency, \textsc{EStream} transforms serialized expert weights offline into the layout
consumed directly by the NPU, eliminating runtime reformatting from the
weight-streaming path. As illustrated in the lower-left of
Figure~\ref{fig:overview}, this transformation reorders the weight data and
groups experts before deployment. It reorganizes
the packed values so that adjacent elements required by an NPU tile are
colocated, and then converts each matrix from row-major to tile-major order. It
further applies the expert permutation and grouping defined above, placing
the gate, up, and down regions of each group in a
contiguous UFS extent. At runtime, vectored reads place these regions directly
into their designated bank locations. Because the loaded bytes already match
the layout expected by the NPU, a completed bank can be published
without CPU-side reformatting or an intermediate copy.

\subsection{Pipelining UFS Streaming and NPU Execution}
\label{sec:design-pipeline}

The bounded arena turns expert streaming into a finite-buffer
producer--consumer pipeline. Without overlap, each bank refill stalls expert
execution, so scheduling determines how much storage latency remains on the
critical path. \textsc{EStream} overlaps UFS loading with NPU execution and
configures grouping, I/O queue depth, and bank count from measured device
behavior.

\noindent\textbf{Constructing the UFS--NPU Pipeline.}
At each sparse layer, the loader begins filling free banks while the NPU executes
non-FFN computation, before routes are available. Because the active set is not
yet known, these initial reads follow the prepacked group order. Once routing
completes, the runtime marks required groups, retires already loaded groups with
no routes, and skips their reads.

Loading and execution then proceed independently. The executor consumes routed
groups whose banks are ready, while the loader refills released banks with later
groups. A load-completion notification publishes a bank only after its data is
ready, and the executor releases it only after its last dependent invocation
completes. Thus, an expert cannot execute before its parameters are ready, and a
bank cannot be overwritten while a queued invocation still references it. Early
loading changes parameter availability without changing routing decisions or
numerical results.

\begin{figure}[!t]
  \centering
  \includegraphics[width=\columnwidth]{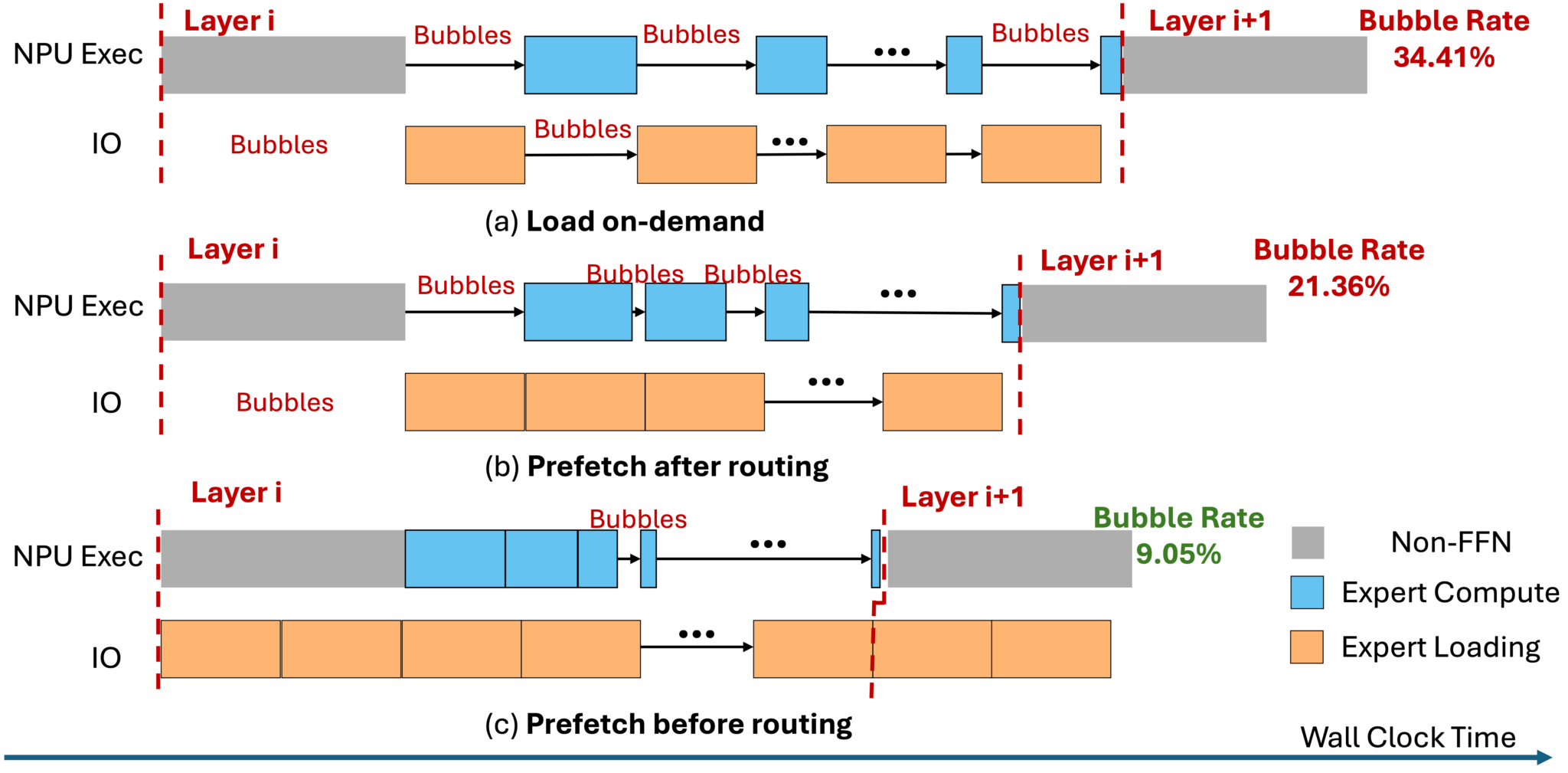}
  \caption{Expert-loading schedules for 1,024-token OLMoE prefill. Earlier
  prefetch increases I/O--compute overlap and lowers bubble rate,
  $1-\max(T_{\mathrm{io}},T_{\mathrm{exec}})/T_{\mathrm{total}}$.}
  \Description{Three timelines compare expert loading and NPU execution.
  On-demand loading alternates orange loading blocks and blue expert-compute
  blocks and has a 34.41 percent bubble rate. Prefetch after routing overlaps
  the two after non-FFN execution and has a 21.36 percent bubble rate. Prefetch
  before routing starts orange loading blocks during gray non-FFN execution
  and reduces the bubble rate to 9.05 percent.}
  \label{fig:pipeline-scheduling}
\end{figure}

Figure~\ref{fig:pipeline-scheduling} compares the three schedules. On-demand
loading places every refill on the critical path and yields a 34.41\% bubble
rate. Prefetching after routing lowers it to 21.36\% but retains the initial fill
bubble. Starting during non-FFN computation reduces it further to 9.05\%.

\begin{figure*}[!t]
  \centering
  \includegraphics[width=\textwidth]{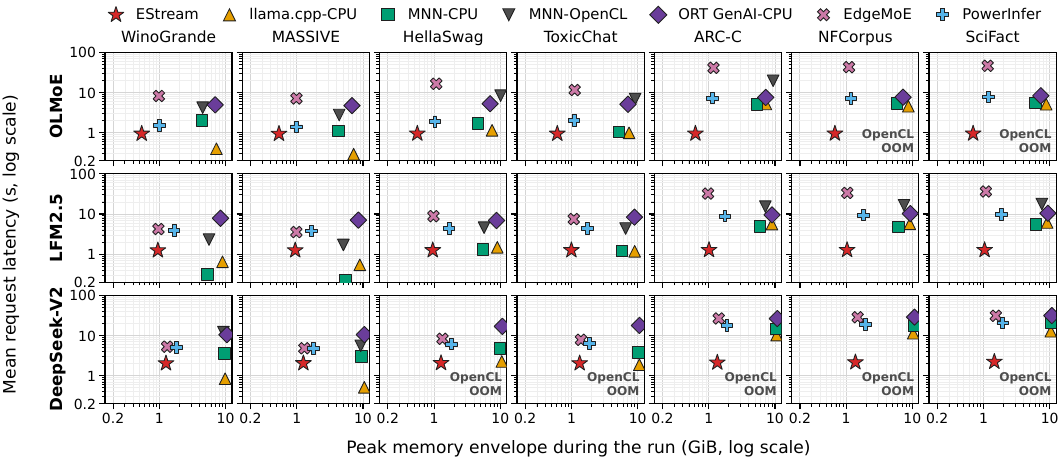}
  \caption{Mean request latency--memory tradeoffs across seven application
  workloads. Rows denote models and columns denote datasets; both axes are
  logarithmic and lower left is better.}
  \Description{A three-by-seven grid of log-scale scatter plots comparing mean
  request latency and memory across seven application datasets. EStream has a
  stable bounded memory footprint and becomes the fastest runtime on the
  longer reasoning and retrieval workloads.}
  \label{fig:dataset32-latency-memory}
\end{figure*}

\noindent\textbf{Automatic Pipeline Configuration.}
\label{sec:auto-pipeline-configuration}
Expert group size $G$, I/O queue depth $Q$, and bank count $B$ jointly determine
pipeline efficiency. Small $G$ fragments reads and increases dispatches,
whereas large $G$ leaves fewer tasks to overlap. Small $Q$ underuses UFS
parallelism, while extra banks only increase residency once neither stage waits
for reuse.

We model configuration as a finite-buffer scheduling problem. For profiled
sparse layer $w$, consider candidate group size $g$, queue depth $q$, and bank
count $b$. Let $k$ index the $K_w(g)$ scheduled groups and let
$\boldsymbol{\rho}^{(g)}_{w,k}$ contain group $k$'s per-expert route counts. Let
$T^{\mathrm{ne}}_w$ denote completion of the layer's non-expert NPU work, and
let $(a_k,d_k)$ and $(s_k,f_k)$ denote the read and execution start--completion
times. With profiled service times $L_{w,k}(g,q)$ and
$X_{w,k}(g,\boldsymbol{\rho}^{(g)}_{w,k})$, the dependencies form the max-plus
recurrence~\cite{baccelli1992synchronization}
\begin{align}
  a_k &= \max(d_{k-q}, f_{k-b}), &
  d_k &= a_k + L_{w,k}(g,q), \nonumber\\
  s_k &= \max(d_k,T^{\mathrm{ne}}_w,f_{k-1}), &
  f_k &= s_k + X_{w,k}(g,\boldsymbol{\rho}^{(g)}_{w,k}),
  \label{eq:pipeline-maxplus}
\end{align}
with $d_j=f_j=0$ for $j\leq0$. In the first maximum, $d_{k-q}$ limits concurrent
reads and $f_{k-b}$ prevents a bank from being reused before its previous NPU
consumer finishes. The second maximum waits for group data, non-expert NPU work,
and prior expert execution. It captures the NPU serialization constraint:
storage reads may overlap NPU computation, but non-expert and expert execution
cannot overlap each other. Thus
$T_{\mathrm{pipe}}(w;g,q,b)=f_{K_w(g)}$, and request time sums this prediction
over its sparse layers.

Let $W_e$ be one prepacked expert's byte size and let $M_{\mathrm{other}}(g)$
include all measured non-arena memory. Under process-memory budget $M_{\max}$,
the target configuration is
\begin{equation}
  \begin{aligned}
  (G^*,Q^*,B^*) \in{}& \arg\min_{G,Q,B}
    \mathbb{E}_{w\sim\mathcal{W}}
    [T_{\mathrm{pipe}}(w;G,Q,B)] \\
  \text{s.t.}\quad & G\in\mathcal{G},\ Q\in\mathcal{Q},\ B\in\mathbb{Z}_{+},\\
  & BGW_e+M_{\mathrm{other}}(G)\leq M_{\max},\quad Q\leq B,
  \end{aligned}
  \label{eq:pipeline-search}
\end{equation}
where $\mathcal{G}$ and $\mathcal{Q}$ are runtime-supported values and
$\mathcal{W}$ is the deployment workload distribution. Exhaustively measuring
every triplet end to end is expensive. We first characterize queue depth once
per device, since it is primarily a property of the UFS/controller path rather
than of model routing. We choose the smallest supported depth that reaches
95\% of peak grouped-read bandwidth for every supported layout,
\begin{equation}
  Q_0=\min\!\left\{q:\
  \min_{g\in\mathcal G}
  \frac{\mathrm{BW}(g,q)}{\max_{q'\in\mathcal Q}\mathrm{BW}(g,q')}
  \geq0.95\right\}.
  \label{eq:pipeline-queue-heuristic}
\end{equation}
On our platform, $Q_0=2$; deeper queues provide no measurable bandwidth gain,
so this value is fixed across models.

Fixing $Q_0$ leaves a budget-coupled $G/B$ problem: a larger group consumes
more memory per bank and simultaneously reduces the number of groups. For each
supported $g$, we first derive the largest feasible bank count
\begin{equation}
  B_{\max}(g)=\min\!\left(
  \left\lfloor\frac{N_e}{g}\right\rfloor,
  \left\lfloor\frac{M_{\max}-M_{\mathrm{other}}(g)}{gW_e}\right\rfloor
  \right).
  \label{eq:pipeline-bank-limit}
\end{equation}
A 1K-token C4 calibration request supplies $T^{\mathrm{ne}}_w$, route vectors,
$L_{w,k}$, and $X_{w,k}$ for each $g$. Since expert groups of the same size
transfer equal numbers of bytes, we aggregate their loading samples within the
request. We also pool the group-independent non-expert times across $g$ to
remove process-level variation. The smallest-$g$ run is memory-monitored and
therefore supplies both its service profile and the non-arena memory
calibration. This reduces profiling to one request per candidate group size.
Equation~\ref{eq:pipeline-maxplus} then evaluates every feasible bank count
without running complete inference. The inner optimization selects
\begin{equation}
  B^*(g)\in\arg\min_{Q_0\leq b\leq B_{\max}(g)}
  \frac{1}{|\mathcal W_{\mathrm{cal}}|}
  \sum_{w\in\mathcal W_{\mathrm{cal}}}
  T_{\mathrm{pipe}}(w;g,Q_0,b),
  \label{eq:pipeline-bank-selection}
\end{equation}
and the outer optimization compares the resulting layouts:
\begin{equation}
  G^*\in\arg\min_{g\in\mathcal G:B_{\max}(g)\geq Q_0}
  \frac{1}{|\mathcal W_{\mathrm{cal}}|}
  \sum_{w\in\mathcal W_{\mathrm{cal}}}
  T_{\mathrm{pipe}}(w;g,Q_0,B^*(g)).
  \label{eq:pipeline-group-selection}
\end{equation}
The emitted $(G^*,Q_0,B^*(G^*))$ is stored with the prepacked model and reused
across requests. This nested search retains the memory coupling between $G$
and $B$ but replaces an exhaustive end-to-end $Q/B/G$ sweep with one short
device profile and inexpensive max-plus evaluation.

\section{Implementation}

\noindent\textbf{Runtime and HTP operators.}
\textsc{EStream} comprises a C++20 host runtime and a low-level HTP
backend extending llama.cpp's open-source Hexagon
support~\cite{ggerganov2023llamacpp}.
It adds approximately 23\,K physical lines of C/C++ excluding inherited code,
tests, and characterization tools. At initialization, the host allocates
reusable graph templates, maps their buffers to cDSP through FastRPC, and
submits descriptors through DSPQueue. On cDSP, scalar workers resolve routes
and addresses, DMA stages tiles, HVX performs data rearrangement and vector
operations, and HMX executes matrix tiles. Our operators cover Q4 embedding,
QKV, MLA, short convolution, routing, and the fused expert path from route-map
construction through weighted scatter-add. 

\noindent\textbf{Numerical representation.}
Expert and large non-expert matrices use weight-only GGUF Q4\_0. OLMoE retains
a Q6\_K output head, while DeepSeek retains FP16 MLA projections. At runtime, DMA
stages packed tiles in VTCM and HVX forms the FP16 tiles consumed by HMX.
Graph interfaces remain FP32 and KV caches use FP16. We apply neither
activation quantization nor approximation-based sparsity.

\noindent\textbf{Expert loading and synchronization.}
The host allocates the expert arena as a UDMABUF-backed DMA-BUF, maps it into
its address space, and exposes the same buffer to the NPU through FastRPC.
Loader threads issue page-aligned \texttt{O\_DIRECT} \texttt{preadv} requests
that place offline-packed expert weights directly into free banks, avoiding an
intermediate staging copy and runtime reformatting. Page population begins
asynchronously during initialization. A mutex-protected placement table and a
condition variable coordinate loading with execution. A loader reserves a bank
before I/O and publishes its expert binding only after the read completes and a
release fence. The executor waits for that binding, retains the bank while its
graph is queued, and releases it only after the NPU completion fence. This
prevents both partially loaded weights and premature bank reuse.

\begin{figure*}[!t]
  \centering
  \includegraphics[width=\textwidth]{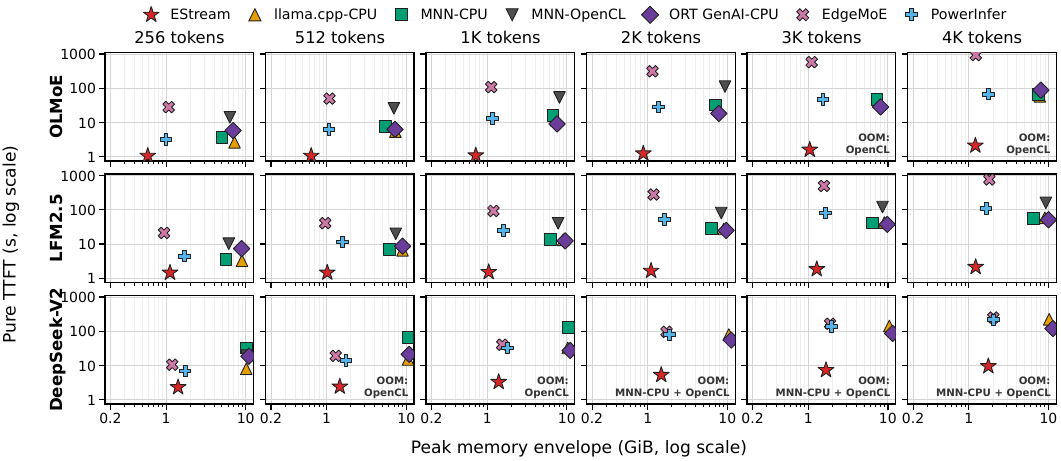}
  \caption{Pure-prefill latency--memory tradeoffs across input lengths on v81.
  Rows denote models and columns denote input lengths; both axes are
  logarithmic and lower left is better.}
  \Description{A three-by-six grid of log-scale scatter plots comparing
  EStream with six baselines. EStream is marked by red stars and generally
  occupies the lower-left region, indicating lower TTFT and memory.}
  \label{fig:prefill-latency-memory}
\end{figure*}

\begin{table}[t]
  \caption{Model structures and deployed payloads. Layers gives total
  (MoE) blocks and $E$ experts per block.}
  \label{tab:eval-models}
  \centering
  \scriptsize
  \setlength{\tabcolsep}{3.0pt}
  \begin{tabular}{@{}lrrrr@{}}
    \toprule
    Model & Layers & $E$ & Non-exp. (GB) & Expert (GB) \\
    \midrule
    OLMoE-1B-7B       & 16 (16) & 64  & 0.30 & 3.62 \\
    LFM2.5-8B-A1B     & 24 (22) & 32  & 0.41 & 4.36 \\
    DeepSeek-V2-Lite  & 27 (26) & 64  & 0.99 & 8.10 \\
    Qwen3-30B-A3B     & 48 (48) & 128 & 0.91 & 16.31 \\
    Mixtral-8$\times$7B & 32 (32) & 8 & 1.04 & 25.37 \\
    \bottomrule
  \end{tabular}
\end{table}

\section{Evaluations}
\label{sec:evaluation}

\subsection{Experiment Settings}

\noindent\textbf{Device and Baselines.}
All on-device measurements use a OnePlus PLK110 phone running Android 16 with
15.1\,GiB of memory. Its Snapdragon 8 Elite Gen 5 platform (SM8850) provides
two prime and six performance Oryon CPU cores, an Adreno GPU, a Hexagon v81
HTP, and UFS 4.1 storage~\cite{qualcomm2025sm8850}. Industrial baselines include
llama.cpp-CPU~\cite{ggerganov2023llamacpp}, MNN-CPU and
MNN-OpenCL~\cite{jiang2020mnn,wang2024mnnllm}, and ONNX Runtime GenAI on
CPU~\cite{microsoft2025ortgenai}. We also reproduce EdgeMoE~\cite{yi2025edgemoe}
and port the parameter-streaming backend from the PowerInfer repository
to the three evaluated models; we refer to this baseline as PowerInfer. 
We tune each supported pair and report
request latency, conservative peak physical memory, and QPT SoC energy under
the protocol specified for each experiment.

\begin{table}[t]
  \caption{Datasets, metrics, and mean prompt lengths.}
  \label{tab:eval-datasets}
  \centering
  \scriptsize
  \setlength{\tabcolsep}{3.5pt}
  \begin{tabular}{@{}lllr@{}}
    \toprule
    Dataset & Task & Metric & Mean tokens \\
    \midrule
    MASSIVE    & Intent      & Accuracy & 16 \\
    WinoGrande & Commonsense & Accuracy & 23 \\
    ToxicChat  & Safety      & F1 & 66 \\
    HellaSwag  & Completion  & Norm. accuracy & 80 \\
    ARC-C      & Science QA  & Accuracy & 368 \\
    NFCorpus   & Retrieval   & nDCG@10 & 376 \\
    SciFact    & Fact check  & F1 & 405 \\
    \bottomrule
  \end{tabular}
\end{table}

\noindent\textbf{Models.}
Our phone experiments cover three models no larger than 16B, namely
OLMoE-1B-7B~\cite{muennighoff2025olmoe},
LFM2.5-8B-A1B~\cite{liquidai2026lfm25}, and
DeepSeek-V2-Lite~\cite{deepseek2024v2}, and test scaling with
Qwen3-30B-A3B~\cite{qwen2025qwen3} and
Mixtral-8$\times$7B~\cite{jiang2024mixtral}. Expert and large non-expert
matrices use GGUF Q4\_0, activations remain FP32, and HMX operands and KV caches
use FP16. OLMoE's output head uses Q6\_K, while DeepSeek's sensitive MLA
projections remain FP16. Table~\ref{tab:eval-models} reports the resulting
measured payloads; DeepSeek's shared experts are resident and counted in its
non-expert payload.

\noindent\textbf{Datasets.}
Our application suite covers commonsense and science reasoning with
WinoGrande~\cite{sakaguchi2020winogrande},
HellaSwag~\cite{zellers2019hellaswag}, and
ARC-Challenge~\cite{clark2018arc}, intent and safety classification with
MASSIVE~\cite{fitzgerald2022massive} and
ToxicChat~\cite{lin2023toxicchat}, and retrieval with
NFCorpus~\cite{boteva2016nfcorpus} and SciFact~\cite{wadden2020fact}.
We share 32 length-stratified examples across systems and tokenize them per
model. Table~\ref{tab:eval-datasets} reports the mean model-token length.
Quality tests additionally use C4 perplexity and bits per byte~\cite{raffel2020t5},
LAMBADA next-word accuracy~\cite{paperno2016lambada}, and BoolQ
accuracy~\cite{clark2019boolq}.

\begin{table}[!t]
  \caption{\textsc{EStream} performance on large MoE models on v81. TTFT is the
  three-run median; memory and QPT energy are measured separately.}
  \label{tab:large-moe-performance}
  \centering
  \scriptsize
  \setlength{\tabcolsep}{3.2pt}
  \begin{tabular}{@{}lrrrr@{}}
    \toprule
    Model & Tokens & Pure TTFT (s) $\downarrow$ &
      Peak mem. (GiB) $\downarrow$ & Tokens/J $\uparrow$ \\
    \midrule
    \multirow{4}{*}{Qwen3-30B-A3B}
      & 1K & 4.838 & 1.461 & 43.59 \\
      & 2K & 5.043 & 1.594 & 62.36 \\
      & 3K & 5.368 & 1.729 & 70.64 \\
      & 4K & 6.836 & 1.866 & 72.18 \\
    \midrule
    \multirow{4}{*}{Mixtral-8$\times$7B}
      & 1K & 7.086 & 2.767 & 23.41 \\
      & 2K & 8.742 & 3.352 & 29.65 \\
      & 3K & 10.971 & 3.917 & 33.17 \\
      & 4K & 13.584 & 4.114 & 35.76 \\
    \bottomrule
  \end{tabular}
\end{table}

\subsection{Performance Evaluation}

\textsc{EStream} advances the latency--memory frontier by bounding expert residency while delivering increasing speedups as
prompt length grows. This advantage persists on application workloads and
enables full-NPU prefill of MoE models with up to 46.7B parameters on a
commercial smartphone.

\begin{figure*}[!t]
  \centering
  \includegraphics[width=\textwidth]{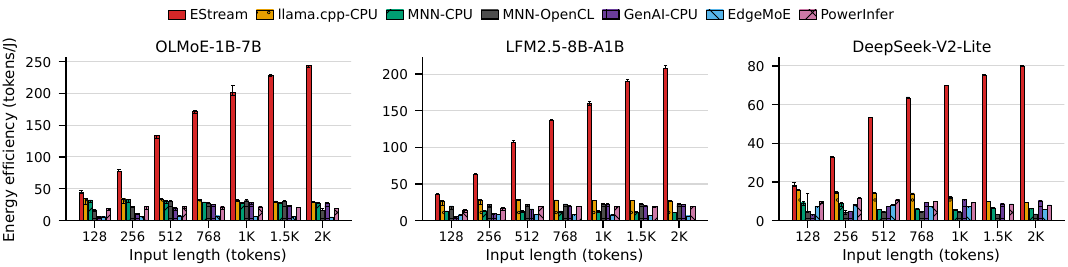}
  \caption{Steady-state prefill energy efficiency across input lengths; higher
  is better. Bars show medians and whiskers show ranges over completed runs.
  The unavailable DeepSeek-V2-Lite EdgeMoE result at 1.5K is omitted.}
  \Description{Three side-by-side grouped bar charts compare the tokens per
  joule of EStream with llama.cpp-CPU, MNN-CPU, MNN-OpenCL, GenAI-CPU,
  EdgeMoE, and PowerInfer on OLMoE, LFM2.5, and DeepSeek-V2-Lite from 128 to
  2,048 input tokens. EStream's efficiency rises substantially with input
  length, while the CPU baselines remain comparatively flat.}
  \label{fig:prefill-energy-efficiency}
\end{figure*}

\noindent\textbf{Performance on Application Workloads.}
Figure~\ref{fig:dataset32-latency-memory} reports mean request latency after one
warm-up and peak memory across seven
datasets whose mean lengths range from 16 to 405 tokens. Peak memory is the larger of
the process-attributed footprint and the decrease in system
\texttt{MemAvailable}. Optimized CPU runtimes
remain faster on MASSIVE and WinoGrande, whose mean lengths are only 16 and 23
tokens, because these requests cannot sufficiently amortize NPU dispatch and
pipeline fill. Nevertheless, \textsc{EStream} uses 5.57--13.43$\times$ less
memory than the latency-leading baseline. The crossover occurs between
ToxicChat at 66 tokens and HellaSwag at 80 tokens. \textsc{EStream} is within
8\% of the fastest baseline on ToxicChat and is 1.03--1.21$\times$ faster on
HellaSwag. On the longer ARC-C, NFCorpus, and SciFact workloads, it is
3.79--5.75$\times$ faster while using 5.91--12.87$\times$ less memory. Across
all 21 model--dataset pairs, \textsc{EStream} outperforms PowerInfer by
1.49--9.29$\times$ and reduces memory by 1.34--1.87$\times$. Even when EdgeMoE
uses up to 3.3\% less memory, its latency remains 2.36--48.45$\times$ higher.

\noindent\textbf{Scaling with Input Length.}
Figure~\ref{fig:prefill-latency-memory} compares prefill TTFT and peak
physical memory for OLMoE-1B-7B, LFM2.5-8B-A1B, and DeepSeek-V2-Lite on C4
prefixes from 256 to 4,096 tokens. Every runtime receives identical token IDs,
and we report the median TTFT of three fresh-process runs after dropping the OS
page cache. \textsc{EStream} completes all configurations within
0.59--1.76\,GiB. Relative to the fastest completed non-offloading baseline at
each point, it improves TTFT by 2.25--27.57$\times$ and reduces physical memory
by 6.45--12.29$\times$. At 4K tokens, its TTFT is 2.12, 2.14, and 9.53 seconds
for OLMoE, LFM2.5, and DeepSeek-V2-Lite, respectively, compared with 58.45,
51.15, and 120.02 seconds for the fastest non-offloading competitors. EdgeMoE
uses up to 16\% less memory on short LFM2.5 and DeepSeek inputs but is at least
4.55$\times$ slower. Against PowerInfer, the strongest offloading baseline,
\textsc{EStream} is 2.96--50.78$\times$ faster and uses
1.16--1.69$\times$ less memory across all 18 configurations.

\noindent\textbf{Scaling to Large MoE Models.}
We further evaluate \textsc{EStream} on Qwen3-30B-A3B and
Mixtral-8$\times$7B, which contain 30.5B and 46.7B parameters while activating
3.3B and 12.9B parameters per token, respectively. To our knowledge,
\textsc{EStream} is the first system to report full-NPU prefill for MoE models
at these scales on a commercial smartphone. As shown in
Table~\ref{tab:large-moe-performance}, Qwen3 sustains 211.7--599.1 tokens/s
within 1.46--1.87\,GiB, while Mixtral sustains 144.5--301.5 tokens/s within
2.77--4.11\,GiB.

\subsection{Energy Efficiency}

Figure~\ref{fig:prefill-energy-efficiency} evaluates steady-state prefill
energy efficiency from 128 to 2,048 input tokens. After loading the model and
one unmeasured warm-up, each runtime processes fixed-length C4 requests with KV
state reset. Qualcomm QPT integrates gross SoC
energy over the measured batch. 

\textsc{EStream} is the most energy-efficient runtime in every completed
model--length setting. Relative to the strongest baseline at each point, it
improves efficiency by 1.19--8.41$\times$, with a geometric mean of
4.08$\times$ across all 21 settings. The gain grows from 1.19--1.34$\times$ at
128 tokens to 7.78--8.41$\times$ at 2K. Table~\ref{tab:large-moe-performance}
extends the result to larger models under a stricter boundary that includes
process launch and initialization. From 1K to 4K, Qwen3-30B-A3B improves from
43.59 to 72.18 tokens/J and Mixtral-8$\times$7B from 23.41 to 35.76 tokens/J.


\begin{figure*}[!t]
  \centering
  \includegraphics[width=\textwidth]{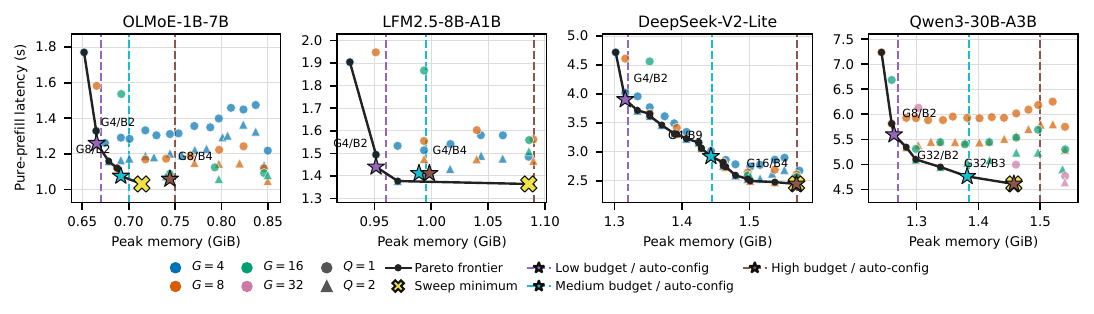}
  \caption{Measured 1K latency--memory tradeoffs across $Q/B/G$
  configurations. Curves show Pareto frontiers, crosses
  mark unconstrained minima, and stars mark auto-configured points under three
  memory budgets.}
  \Description{Four scatter plots show pure-prefill latency against peak
  memory for OLMoE, LFM2.5, DeepSeek-V2-Lite, and Qwen3-30B-A3B. Each plot
  overlays a measured Pareto frontier, the global sweep minimum, three memory
  budgets, and the configuration selected automatically under each budget.}
  \label{fig:qbg-pareto}
\end{figure*}

\begin{figure}[t]
  \centering
  \includegraphics[width=\columnwidth]{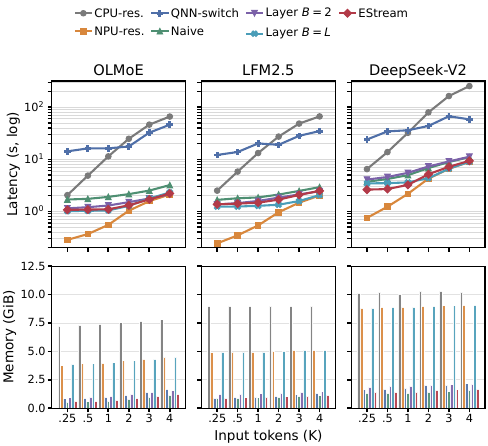}
  \caption{System ablation across compute and expert-loading strategies.
  The top row shows pure-prefill latency, and the bottom row shows peak
  physical memory.}
  \Description{A two-by-three plot compares OLMoE, LFM2.5, and
  DeepSeek-V2-Lite. The top row uses lines to show pure-prefill latency, while
  the bottom row uses grouped bars to show peak physical memory. Each column
  compares fully resident CPU and NPU execution, per-expert QNN context
  streaming, naive NPU streaming with Q equals 1, G equals 1, and B equals 3,
  two cross-layer loading strategies, and the complete EStream system from 256
  to 4,096 input tokens.}
  \label{fig:streaming-ablation}
\end{figure}

\subsection{Model Accuracy}
\label{sec:evaluation-accuracy}

\begin{table}[!t]
  \caption{Unified Q4 quality on the same 200 source examples per dataset.
  C4 is word PPL ($\downarrow$); other values are percentages ($\uparrow$).
  For paired families, bold compares Host Dense against \textsc{EStream} MoE.}
  \label{tab:unified-quality}
  \centering
  \scriptsize
  \setlength{\tabcolsep}{1.2pt}
  \renewcommand{\arraystretch}{0.94}
  \resizebox{\columnwidth}{!}{%
  \begin{tabular}{@{}llrrrrrrr@{}}
    \toprule
    Checkpoint & Exec. & C4 $\downarrow$ & Wino. $\uparrow$
      & ARC-C $\uparrow$ & LAMB. $\uparrow$ & Hella. $\uparrow$
      & BoolQ $\uparrow$ & SciFact $\uparrow$ \\
    \midrule
    OLMo-1B & Host & 33.52 & 65.00 & 28.00 & 64.50
      & 65.50 & 63.50 & 18.59 \\
    \midrule
    \multirow{2}{*}{OLMoE-1B-7B} & Host & 29.15 & 68.00
      & 63.50 & 75.00 & 78.50 & 76.50 & 20.97 \\
      & \textsc{EStream} & \textbf{29.15} & \textbf{68.50}
      & \textbf{63.50} & \textbf{74.00} & \textbf{77.50}
      & \textbf{76.50} & \textbf{22.12} \\
    \midrule
    \midrule
    LFM2.5-1.2B & Host & 51.44 & 59.00 & 74.00 & 49.50
      & 63.50 & \textbf{75.00} & \textbf{20.17} \\
    \midrule
    \multirow{2}{*}{LFM2.5-8B-A1B} & Host & 36.51 & 67.00
      & 85.00 & 59.50 & 73.50 & 72.50 & 17.82 \\
      & \textsc{EStream} & \textbf{36.48} & \textbf{67.50}
      & \textbf{85.00} & \textbf{59.50} & \textbf{74.50}
      & 74.00 & 17.27 \\
    \midrule
    \midrule
    Qwen3-4B & Host & 50.38 & \textbf{72.00} & 89.00 & 63.50
      & 68.00 & 85.00 & 55.21 \\
    \midrule
    \multirow{2}{*}{Qwen3-30B-A3B} & Host & 31.41 & 72.50
      & 95.00 & 75.50 & 78.50 & 91.50 & 61.64 \\
      & \textsc{EStream} & \textbf{33.14} & 71.50
      & \textbf{94.50} & \textbf{72.50} & \textbf{76.50}
      & \textbf{90.50} & \textbf{60.00} \\
    \midrule
    \midrule
    \multirow{2}{*}{DeepSeek-V2-Lite} & Host & 37.50 & 76.50
      & 74.50 & 74.00 & 79.00 & 85.50 & 18.18 \\
      & \textsc{EStream} & 37.47 & 76.00
      & 71.50 & 74.50 & 79.50 & 86.00 & 18.89 \\
    \bottomrule
  \end{tabular}%
  }
\end{table}

\textsc{EStream} retains the deployed Q4 weights and routing decisions but
changes finite-precision operation order. Its fused NPU path tiles projections
differently, executes expert groups as their weights become ready, and
scatter-adds routes in a different sequence. Table~\ref{tab:unified-quality}
compares each MoE checkpoint on \textsc{EStream} and the Host reference using
the same 200 examples and model-specific tokenization.

The resulting quality differences are small. OLMoE and LFM2.5 change C4 PPL
by less than 0.1\% and every task score by at most 1.5 points. DeepSeek has
similarly stable PPL and changes five of six task scores by at most 0.71 points,
with a 3.00-point ARC-C exception. Qwen3 has the largest deviation: a 5.51\%
relative PPL increase and at most a 3.00-point task-score change. Reordering
computation for streaming full-NPU execution therefore generally preserves
end-to-end model quality, although Qwen3 remains the least numerically aligned
path.

We additionally compare the MoE checkpoints deployed by \textsc{EStream} with
same-family dense checkpoints of comparable active-parameter scale executed on
the host CPU. Across the three model families, the MoE deployments achieve
better results on 18 of the 21 reported metrics. They reduce C4 PPL by
13.0--34.2\%. OLMoE improves all six downstream-task scores, while LFM2.5 and
Qwen3 improve four and five, respectively. This comparison demonstrates the
practical advantage of deploying these higher-capacity MoE models.

\subsection{Pipeline Configuration and Auto-Tuning}
\label{sec:evaluation-pipeline-configuration}

Figure~\ref{fig:qbg-pareto} shows how the pipeline parameters shape the
latency--memory operating point and whether the configurator in
Section~\ref{sec:auto-pipeline-configuration} can find a good point without an
end-to-end sweep. For an exact 1,024-token C4 prefix, we exhaustively evaluate
all production-supported $Q/B/G$ configurations: 25 for LFM2.5 and 53 for each
of the other three models, totaling 184. 
The sweep exposes strong coupling among the three parameters.
At OLMoE $G=8,B=3$, increasing $Q$ from 1 to 2 reduces latency by 11.8\% by
exposing the available UFS parallelism. With $G=8,Q=2$, increasing $B$ from 2
to 3 further reduces latency from 1.075 to 1.030\,s, but retaining all eight
banks takes 1.048\,s while using 16.0\% more memory than the three-bank point.
Thus, banks help only until the producer can keep the NPU supplied. Group size
changes both request granularity and the number of feasible banks: relative to
the best feasible fixed-$G=8$ configurations, searching $G$ reduces latency by
4.8\% at the medium DeepSeek budget and by 14.5\% at the high Qwen3 budget.
No single parameter can therefore be maximized or fixed independently.

The auto-configurator closely tracks the sweep oracle. Across the 12
model--budget pairs, it exactly matches 9 constrained minima. All selected
points satisfy their measured budgets, with zero median regret and 3.45\%
worst-case regret. The three misses occur on one OLMoE budget and two LFM2.5
budgets, where bank-dependent synchronization costs depart from the profiled
service model. These results show that lightweight hardware profiles recover
near-Pareto configurations without an exhaustive deployment-time search. With
one service-profile request per group size and memory monitoring folded into
one request, profiling and selection take 5.7--22.4\,s. This is
34.2--76.7$\times$ faster than the 302--1,685\,s exhaustive sweeps under the
same active process-time accounting. The compact profiles are reusable across
memory budgets, after which selecting a new configuration takes only
0.34--2.79\,ms.

\subsection{System Ablation}
\label{sec:evaluation-system-ablation}

\noindent\textbf{Full-NPU execution and expert virtualization.}
To measure their combined effect, Figure~\ref{fig:streaming-ablation} compares
\textsc{EStream} with two fully resident references. CPU-resident uses the
optimized llama.cpp CPU path. NPU-resident uses the same EStream NPU execution path, 
but preloads every model parameters. It therefore isolates the overhead of expert streaming. 
Compared with CPU-resident,
\textsc{EStream} is 1.83--30.14$\times$ faster and uses 6.05--12.29$\times$
less memory. The I/O-free NPU-resident reference is 1.04--5.63$\times$ faster,
but consumes 3.73--6.36$\times$ more memory. Longer input creates more overlapping windows. 
At 4K, \textsc{EStream} incurs only 4.3--23.0\% higher latency than resident
execution while using 3.73--5.41$\times$ less memory,
and the LFM2.5 trace shows that 94.9\% of UFS loading overlaps NPU execution.

\noindent\textbf{Topology-invariant graph sharing.}
As shown in Figure~\ref{fig:shared-expert-context-comparison}, QNN-switch uses
the same full-NPU non-expert path and approximately the same
expert-residency budget as \textsc{EStream}. However, it prepares a fixed QNN
graph for every expert and loads the selected expert contexts at runtime.
This isolates the benefit of sharing one expert graph and binding its
parameters dynamically. Despite using only 1.00--1.41$\times$ the memory of
\textsc{EStream}, QNN-switch is 6.27--21.02$\times$ slower because context
loading and repeated graph invocation remain on the critical path.

\noindent\textbf{Fine-grained I/O--execution pipeline.}
Naive streaming retains the shared graph but uses $Q=1$, $G=1$, and $B=3$.
Automatic grouping and buffering make \textsc{EStream} 1.18--1.73$\times$
faster for only 2.7--14.1\% more memory. We further compare two coarse
cross-layer schedules with $G=E$ and $Q=2$. Layer-$B{=}2$ alternates two
layer-sized arenas, whereas Layer-$B{=}L$ keeps one arena per MoE layer.
\textsc{EStream} outperforms Layer-$B{=}2$ in 15 of 18 cases by
1.02--1.71$\times$ while using 1.26--1.58$\times$ less memory.
Layer-$B{=}L$ removes reuse stalls and can reduce latency by up to 23.6\%
(1.31$\times$), but requires
3.73--6.61$\times$ more memory. Thus, coarse layer streaming either stalls at
layer boundaries or abandons bounded residency, while expert-group pipelining
provides a better latency--memory balance.

\section{Related Work}

\noindent\textbf{LLM parameter offloading.}
Prior systems move weights across memory and storage tiers to enlarge effective
model capacity. FlexGen and LLM in a Flash optimize device placement and flash
access for dense models~\cite{sheng2023flexgen,alizadeh2023llmflash}. MoE
systems cache or prefetch experts~\cite{eliseev2023moeoffloading,
xue2024moeinfinity,song2024promoe,yu2026finemoe}, reduce transfer precision
~\cite{yi2025edgemoe,tang2024hobbit,wang2025d2moe}, or pipeline storage with
CPU/GPU execution~\cite{cao2025moelightning,su2026zeroprefill}. These designs
primarily target decode locality, rely on CPU/GPU computation, or alter
precision. None supports full-NPU MoE prefill acceleration on mobile devices. 
PowerInfer-2 comes closest by overlapping UFS reads with NPU-centric prefill, 
but uses precompiled graph variants and
lacks a public mobile implementation~\cite{xue2024powerinfer2}. \textsc{EStream}
for the first time realized efficient parameter streaming for dynamic mobile NPU execution
and achieves the best performance.

\noindent\textbf{On-device LLM prefill acceleration.}
Prior work accelerates on-device prefill through computation reuse, model
specialization, and heterogeneous execution. AttnCache reuses approximate
attention maps, while PRISM prunes low-ranked candidates and streams model
layers~\cite{song2025attncache,zhou2026prism}. MNN-LLM and Transformer-Lite
optimize quantized CPU/GPU execution and memory management, whereas MobileMoE
co-designs compact MoE architectures for mobile constraints
~\cite{wang2024mnnllm,li2024transformerlite,chen2026mobilemoe}. llm.npu and
HeteroInfer partition dense LLM computation across mobile processors, while
KTransformers applies asynchronous CPU/GPU execution to MoEs~\cite{xu2025llmnpu,chen2025heteroinfer,chen2025ktransformers}.
NPUMoE is closest in workload and achieves strong speedups using
capacity-tiered, grouped expert graphs on Apple NPUs
~\cite{benazir2026npumoe}. However, its expert graphs have precompiled
capacities, routing and aggregation remain on the CPU, and all model weights
are assumed resident. It therefore cannot bind runtime route extents and
streamed expert addresses within reusable NPU execution. \textsc{EStream}
provides this dynamism while bounding expert residency on smartphones.

\noindent\textbf{Programmable mobile-NPU execution.}
Most mobile inference engines access NPUs through vendor-compiled operator
graphs. Recent work instead exposes programmable execution below this
abstraction. Hao et al. build a FastRPC-connected Hexagon operator library
that directly coordinates HMX, HVX, DMA, and on-chip memory for quantized LLM
inference~\cite{hao2026mobilenpu}. Hexagon-MLIR compiles Triton kernels and
PyTorch graphs into Hexagon binaries, automating fusion, TCM tiling, HVX
vectorization, multithreading, and asynchronous DMA~\cite{absar2026hexagonmlir}.
llada.cpp builds direct Hexagon kernels for diffusion LLMs~\cite{wang2026lladacpp}. 
These systems establish the benefits of programming mobile NPUs below graph operators,
yet do not support the dynamic MoE prefill nor efficient memory management for MoEs. 

\section{Discussion}

\noindent\textbf{More aggressive quantization.}
\textsc{EStream} currently uses weight-only Q4\_0 without activation
quantization. More aggressive low-bit and expert-wise mixed-precision schemes
are largely orthogonal to our design. EdgeMoE assigns expert bit widths offline according to accuracy
sensitivity. D$^2$MoE instead selects nested bit-width representations at
runtime~\cite{yi2025edgemoe,wang2025d2moe}. Integrating such schemes would
shrink streamed expert groups, reduce UFS traffic, and fit more groups within a
fixed memory budget. The latency gain, however, need not scale linearly with
the byte reduction. The NPU must unpack the encoded weights, apply their
quantization parameters, and form HMX-consumable operands. Lower or mixed
precision can therefore lengthen the HVX stage and disturb the
DMA--HVX--HMX pipeline. A quantized extension should jointly optimize quality,
I/O latency, and NPU dequantization cost.

\noindent\textbf{MoE decode.}
\textsc{EStream} targets prefill-only workloads and does not optimize
autoregressive decode. A decode step supplies only one routes to each
selected expert, leaving little matrix work to saturate HMX or hide a
storage access. 
Extending \textsc{EStream} to MoE decode will require a decode-specific policy 
that retains
experts across steps, predicts upcoming routes, and potentially batches or
speculatively verifies tokens to create a larger NPU workload. These techniques
can reuse its memory management and NPU execution, but require a
separate optimization of cache hit rate, tail TPOT, memory, and energy. We
leave this co-design to future work.

\noindent\textbf{Portability across NPUs.}
Our implementation is based on Qualcomm Hexagon, but the design does not depend on
a particular number of vector or matrix units. Porting \textsc{EStream} to other
platforms only requires an programmable NPU interfaces for vector and matrix computation 
and asynchronous data movement. Topology-invariant graph sharing
applies directly when experts within a layer use the same computational
structure. Models with heterogeneous expert dimensions or additional
expert-specific operators may instead require multiple shared graph templates.

\section{Conclusion}

On-device MoE prefill is constrained by a fundamental mismatch between
input-dependent expert execution, static NPU abstractions, and the
request-level densification of expert working sets. This paper presented
\textsc{EStream}, which separates reusable expert computation topology from
runtime route and parameter bindings, streams expert weights through a bounded
NPU-addressable arena, and overlaps UFS loading with full-NPU prefill
execution. On a commercial Snapdragon smartphone, \textsc{EStream} consistently
achieves a better latency--memory operating point across structurally distinct
MoE models and input lengths from 256 to 4,096 tokens. It accelerates resident
CPU and QNN context-switching baselines by up to 30.14$\times$ and
21.02$\times$, respectively. Compared with a same-backend, I/O-free resident
reference, it requires 3.73--6.36$\times$ less peak physical memory while
incurring only 4.3--23.0\% higher latency at 4K. \textsc{EStream} also provides
the highest energy efficiency in every completed model--length setting.
More broadly, these results show that deployable MoE capacity need not be
limited by simultaneously resident expert memory when programmable NPU
execution is co-designed with storage streaming.

\bibliographystyle{ACM-Reference-Format}
\bibliography{references}

\end{document}